\documentclass[aps,pra,reprint,superscriptaddress,
nofootinbib]{revtex4-2}

\usepackage{comment}
\usepackage{graphicx}% Include figure files
\usepackage{dcolumn}% Align table columns on decimal point
\newcolumntype{d}{D{.}{.}{-1}}
\usepackage{bm}% bold math
\usepackage{booktabs}
\usepackage{amsmath} % for aligned environment
\usepackage[dvipsnames]{xcolor}
\usepackage[normalem]{ulem}
\usepackage{soul}
\usepackage{xcolor}

\newcommand{\gl}[1]{{\color{black}#1}}
\newcommand{\ha}[1]{{\color{black}#1}}

\begin{document}
%\preprint{APS/123-QED}
\title{ Gaussian quantum reservoir computing with a hybrid cavity magnomechanical system}
\author{Hajar Assil}
\email{assil.hajar@etu.uae.ac.ma}
\affiliation{Laboratory of R\&D in Engineering Sciences, Faculty of Sciences and Techniques Al-Hoceima, Abdelmalek Essaadi University, Tetouan, Morocco}
\affiliation{%
 Institute for Cross-Disciplinary Physics and Complex Systems (IFISC) UIB-CSIC, Campus Universitat Illes Balears, 07122 Palma de Mallorca, Spain.
}%

%\author{Khadija El Anouz}
%\affiliation{Laboratory of R\&D in Engineering Sciences, Faculty of Sciences and Techniques Al-Hoceima, Abdelmalek Essaadi University, Tetouan, Morocco}

\author{Abderrahim El Allati}
\affiliation{Laboratory of R\&D in Engineering Sciences, Faculty of Sciences and Techniques Al-Hoceima, Abdelmalek Essaadi University, Tetouan, Morocco}

\author{Gian Luca Giorgi}%
\email{gianluca@ifisc.uib-csic.es}
\affiliation{%
 Institute for Cross-Disciplinary Physics and Complex Systems (IFISC) UIB-CSIC, Campus Universitat Illes Balears, 07122 Palma de Mallorca, Spain.
}%

\date{\today}

\begin{abstract}
We propose a quantum reservoir computing framework based on a hybrid cavity magnomechanical system in the linearized Gaussian regime. The reservoir combines microwave-cavity, magnon, and mechanical degrees of freedom, and is extended by an auxiliary cavity acting as an input port, with time-dependent signals encoded in its detuning. The covariance matrix of the quadrature fluctuations provides the features for a trained linear readout. Using linear-memory, nonlinear-memory, and parity-check benchmarks, we find strong temporal memory together with more limited nonlinear processing, whose balance is controlled by the reservoir evolution time, and we show that intermode correlations substantially enhance the information accessible to the readout. The same architecture reconstructs a time-dependent signal encoded in the auxiliary-cavity detuning, with an accuracy governed by the interplay between the internal couplings and the encoding strength, and robust against Gaussian detuning noise. Accounting for finite measurement statistics reveals a trade-off between encoding strength and the precision of the covariance estimation, so that the optimal encoding depends on the available measurement budget. These results establish hybrid cavity magnomechanical systems as a promising platform for continuous-variable quantum reservoir computing with potential applications in signal probing.

\end{abstract}

\maketitle

\section{Introduction}
Reservoir computing (RC) has emerged as a powerful paradigm for processing
temporal information. In contrast to conventional recurrent neural networks,
where all internal weights are trained, RC leaves the dynamics of a fixed
nonlinear system, the reservoir, untouched, and trains only a linear readout
acting on the reservoir response~\cite{Jaeger2001, Maass2002, JaegerHaas2004, Lukosevicius2009}. This drastically reduces the training cost
while retaining the ability to perform complex temporal transformations.
More broadly, a wide range of physical systems has been shown to
act as effective reservoirs, giving rise to the field of physical
reservoir computing~\cite{Tanaka2019, Nakajima2020}, in which the
computation is delegated to the intrinsic dynamics of a hardware
substrate rather than to a digitally simulated network.
When the dynamics of a quantum system realize the reservoir, the framework is
known as quantum reservoir computing (QRC)~\cite{Fujii_2017, Mujal2021}. Since
its introduction, QRC has been applied to temporal information processing, to
the characterization of quantum states, and to a variety of classical and
quantum learning tasks~\cite{Ghosh2019, Mujal2023, Yan2024}, exploiting the
dynamics of the physical substrate as a computational resource with a
comparatively simple training procedure. Beyond numerical studies, QRC has by
now been demonstrated experimentally on several platforms, including
nuclear-spin ensembles~\cite{Negoro2018, Hou2026}, superconducting processors
in both gate-based~\cite{Chen2020, Suzuki2022, Hu2024} and analog
microwave~\cite{Senanian2024, Carles2025} implementations, neutral-atom arrays
with up to one hundred qubits~\cite{Kornjaca2024}, and photonic systems,
ranging from multimode squeezed
light~\cite{Paparelle2026} and integrated circuits with quantum
memristors~\cite{Selimovic2025} or multiphoton inputs~\cite{DiBartolo2026}
to Gaussian boson samplers with hundreds of modes~\cite{Cimini2026}. These experiments confirm that the fixed
dynamics of a physical quantum system can be harnessed for temporal processing
under realistic noise and finite sampling.

Among the possible substrates, continuous-variable Gaussian systems have
recently attracted particular attention. Their state is fully specified by
first and second moments---the mean quadratures and the covariance
matrix---which provide a natural and experimentally accessible set of features
for encoding and manipulating temporal information. Gaussian states have been
shown to constitute a universal and versatile resource for reservoir
computing~\cite{Nokkala2021}, and the role of quantum resources such as
squeezing has been investigated in this context~\cite{GarciaBeni2024,LabayMora2024}. RC has
moreover been demonstrated experimentally with multimode squeezed optical
states~\cite{Paparelle2026}, on platforms designed for real-time
continuous-variable processing~\cite{GarciaBeni2023}. These results motivate the
search for Gaussian reservoirs in other physical platforms, and in particular in
hybrid ones, where modes of different physical nature interact and give rise to
rich dynamical responses.

Hybrid cavity magnomechanical systems are a promising candidate in this
respect. Building on the strong coupling between magnons and microwave cavity
photons~\cite{Huebl2013, Tabuchi2014, Zhang2014}, these platforms combine
photonic, magnonic, and mechanical degrees of freedom: microwave-cavity photons
couple to the collective spin excitations (magnons) of a ferromagnetic crystal
such as yttrium iron garnet, while the magnons couple to a mechanical resonator
through the magnetostrictive interaction~\cite{Kittel1958, Zhang2016, Li2018}.
The coexistence of these degrees of freedom, interacting across very different
frequency and dissipation scales and driven into a linearized Gaussian regime,
produces exactly the kind of multimode correlated dynamics that a Gaussian
reservoir can exploit~\cite{Rameshti2022, Zuo2024}. Their potential as
computational substrates for QRC, however, has remained largely unexplored.

In this work, we propose and analyze a hybrid cavity magnomechanical system as a Gaussian quantum reservoir. To inject information without disturbing the intrinsic interactions of the reservoir, we extend the system with an auxiliary cavity that acts as an input port. A time-dependent signal is encoded in the detuning of this auxiliary mode, which is then processed by the coupled dynamics of the auxiliary cavity, microwave cavity, magnon, and mechanical modes. The mode parameters remain fixed. In the linearized regime, the dynamics is Gaussian. We build the reservoir features from the quadrature fluctuations and correlations contained in the covariance matrix. We assess the computational capabilities of the platform using standard benchmarks, such as linear and nonlinear memory and the parity-check task. We characterize the trade-off between memory and nonlinearity as a function of the reservoir evolution time and analyze how the choice of accessible observables shapes performance. Our results show that inter-mode correlations play an essential role.

Finally, because information is injected through a physical parameter of the
system, the auxiliary-cavity detuning, the same architecture naturally extends
itself to applications beyond benchmark computation. Reconstructing a
time-dependent signal encoded in a system parameter directly from the reservoir
response points to the use of the reservoir as a probe of externally encoded signals~\cite{Angelatos2021, Senanian2024}, connecting to recent learning-based approaches to the inference of quantum-state properties, both theoretical~\cite{Assil2025, Assil2026} and experimental~\cite{Suprano2024, Zia2025}. We illustrate this possibility here with a
proof-of-principle example.

\section{Physical Model}

\subsection{Hybrid cavity magnomechanical system}

We consider a hybrid cavity magnomechanical system composed of a microwave
cavity, a magnon mode supported by a yttrium iron garnet (YIG) sphere, and a
mechanical resonator~\cite{Zhang2016}. The cavity and magnon
modes exchange excitations through the magnetic dipole interaction with strength
$G_{mc}$, while the magnetostrictive deformation of the YIG sphere couples the
magnon to the mechanical resonator with single-magnon coupling $g_{ms}$. The
magnon mode is coherently driven by an external microwave field of amplitude
$\Omega$ and frequency $\omega_d$, which enhances the magnomechanical
interaction and is taken as the reference frequency in the following. Under this
continuous driving, the coupled dynamics is Gaussian and forms the basis of the
reservoir. The Hamiltonian reads
\begin{equation}
\begin{aligned}
\mathcal{H}/\hbar
=&\ \omega_c c^{\dagger}c + \omega_m m^{\dagger}m
+ \frac{\omega_s}{2}\left(q^2+p^2\right) \\
& + g_{ms}m^{\dagger}mq
+ G_{mc}\left(c+c^{\dagger}\right)\left(m+m^{\dagger}\right) \\
& + i\Omega\left(m^{\dagger}e^{-i\omega_dt}-me^{i\omega_dt}\right),
\end{aligned}
\label{ham1}
\end{equation}
where $c$ ($c^\dagger$) and $m$ ($m^\dagger$) are the bosonic annihilation
(creation) operators of the cavity and magnon modes, obeying $[O,O^\dagger]=1$
($O=c,m$), while $q$ and $p$ are the dimensionless position and momentum
quadratures of the mechanical resonator, with $[q,p]=i$. The frequencies
$\omega_c$, $\omega_m$, and $\omega_s$ denote the cavity, magnon, and mechanical
resonances, respectively.

To remove the explicit time dependence of the drive, we move to a frame rotating
at $\omega_d$, in which the cavity and magnon frequencies are replaced by the
detunings $\Delta_c=\omega_c-\omega_d$ and $\Delta_m=\omega_m-\omega_d$. Since
the resonance frequencies far exceed the couplings and dissipation rates, we
adopt the rotating-wave approximation, which casts the cavity--magnon
interaction in the beam-splitter form $G_{mc}(cm^\dagger+c^\dagger m)$. The
Hamiltonian then becomes
\begin{equation}
\begin{aligned}
\mathcal{H}/\hbar =\ & \Delta_c c^{\dagger} c+\Delta_m m^{\dagger} m
+\frac{\omega_s}{2}\left(q^2+p^2\right) \\
& +g_{ms} m^{\dagger} m q+G_{mc}\left(c m^{\dagger}+c^{\dagger} m\right)
+i \Omega\left(m^{\dagger}-m\right).
\end{aligned}
\label{H1}
\end{equation}

\subsection{Extended hybrid cavity magnomechanical system}

To use the hybrid system as a reservoir, we require an interface through which an
external signal can be injected without altering the intrinsic
cavity--magnon--mechanical interactions. To this end we couple an auxiliary
cavity mode $a$, with $[a,a^\dagger]=1$, to the microwave cavity through a
beam-splitter interaction of strength $J$, so that $a$ acts as the input port of
the reservoir. The extended Hamiltonian reads
\begin{equation}
\begin{aligned}
\mathcal{H}_E/\hbar &= \Delta_c c^\dagger c + \Delta_a a^\dagger a
+ \Delta_m m^\dagger m + \frac{\omega_s}{2}(q^2+p^2) \\
& + g_{ms}m^\dagger m\, q + G_{mc}\left(c m^{\dagger}+c^{\dagger} m\right) \\
& + J(c^\dagger a + ca^\dagger) + i\Omega(m^\dagger - m),
\end{aligned}
\label{ham_ex}
\end{equation}
where $\Delta_a=\omega_a-\omega_d$ is the detuning of the auxiliary cavity from
the drive. Information is injected by modulating this detuning at each step $k$,
\begin{equation}
\Delta_a^{(k)} = \Delta_{a,0} + \epsilon_\Delta s_k,
\label{eq:encoding}
\end{equation}
with $\Delta_{a,0}$ the operating detuning, $s_k$ the input sequence, and
$\epsilon_\Delta$ the encoding strength. Because the input enters only through $a$, the
internal interactions of the reservoir are left untouched, while $J$ allows the
encoded information to propagate throughout the hybrid system.

Each mode is additionally coupled to its
environment~\cite{inbook}: the
microwave, auxiliary, and magnon modes decay at rates $\kappa_c$, $\kappa_a$,
and $\kappa_m$, respectively, and the mechanical resonator at rate $\gamma_s$.
The corresponding quantum noise is introduced explicitly through the quantum
Langevin equations of the following section.

Unless otherwise stated, the simulations are performed using the
parameters summarized in Table~\ref{tab:parameters}. The values of
the cavity--magnon--mechanical subsystem are chosen within experimentally relevant regimes of cavity magnomechanical platforms~\cite{Zhang2016,naimy2026enhancedmultiparameterquantumestimation}, while the auxiliary-cavity parameters define the operating point adopted in this work.

\begin{table}[h]
\centering
\caption{Parameters of the hybrid cavity magnomechanical reservoir used throughout
this work, unless otherwise stated. Frequencies and rates are given as
$X/2\pi$; the auxiliary-cavity operating detuning $\Delta_{a,0}$ and the
encoding strength $\epsilon_\Delta$ assume the values specified here unless differently indicated.}
\label{tab:parameters}
\begin{ruledtabular}
\begin{tabular}{lcc}
Parameter & Symbol & Value \\
\hline
Microwave-cavity frequency
& $\omega_c/2\pi$
& $10~\mathrm{GHz}$ \\

Auxiliary-cavity frequency
& $\omega_a/2\pi$
& $10~\mathrm{GHz}$ \\

Magnon frequency
& $\omega_m/2\pi$
& $10~\mathrm{GHz}$ \\

Mechanical frequency
& $\omega_s/2\pi$
& $10~\mathrm{MHz}$ \\

Microwave-cavity decay rate
& $\kappa_c/2\pi$
& $6~\mathrm{MHz}$ \\

Auxiliary-cavity decay rate
& $\kappa_a/2\pi$
& $1~\mathrm{MHz}$ \\

Magnon decay rate
& $\kappa_m/2\pi$
& $1~\mathrm{MHz}$ \\

Mechanical damping rate
& $\gamma_s/2\pi$
& $100~\mathrm{Hz}$ \\

Cavity--magnon coupling
& $G_{mc}/2\pi$
& $6~\mathrm{MHz}$ \\

Effective magnomechanical coupling
& $G_{ms}/2\pi$
& $6~\mathrm{MHz}$ \\

Auxiliary--cavity coupling
& $J/2\pi$
& $5~\mathrm{MHz}$ \\

Encoding strength
& $\epsilon_\Delta/2\pi$
& $0.2~\mathrm{MHz}$ \\

Temperature
& $T$
& $50~\mathrm{mK}$ \\
\end{tabular}
\end{ruledtabular}
\end{table}

\begin{figure}[ht]
  \centering
  \includegraphics[width=0.5\textwidth]{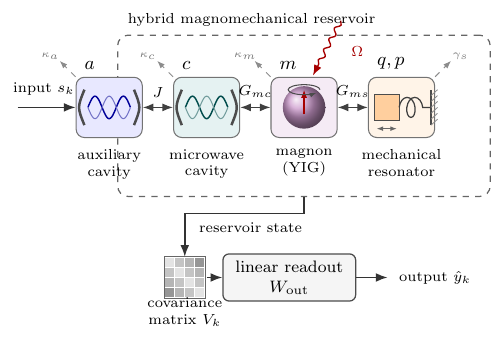}
  \caption{Schematic of the cavity magnomechanical quantum reservoir. An input
sequence $s_k$ is injected by modulating the detuning $\Delta_a^{(k)}$ of an
auxiliary cavity $a$, which is coupled with strength $J$ to the microwave
cavity $c$ of a hybrid magnomechanical system. The cavity exchanges
excitations with the magnon mode $m$ of a YIG sphere at rate $G_{mc}$, and
the magnon couples to a mechanical resonator ($q,p$) through the
magnetostrictive interaction, with effective coupling $G_{ms}$; the magnon is
coherently driven with Rabi frequency $\Omega$, and the modes are damped at
rates $\kappa_a$, $\kappa_c$, $\kappa_m$, and $\gamma_s$. After each input
step, the covariance matrix $V_k$ of the quadrature fluctuations provides the
reservoir features, from which a trained linear readout $W_{\rm out}$ produces
the output $\hat y_k$.}
\label{fig:scheme}
  \label{fig:scheme}
\end{figure}

\subsection{System dynamics}

The evolution of the hybrid system is described by the set of quantum Langevin equations \cite{Li2018, inbook, PhysRevLett.98.030405,PhysRevLett.46.1}

\begin{align}
\dot c &= -(\kappa_c+i\Delta_c)\,c - iG_{mc}\,m - iJ\,a
          + \sqrt{2\kappa_c}\,c^{\rm in}, \\
\dot a &= -\big(\kappa_a+i\Delta_a^{(k)}\big)\,a - iJ\,c
          + \sqrt{2\kappa_a}\,a^{\rm in}, \\
\dot m &= -(\kappa_m+i\Delta_m)\,m - iG_{mc}\,c - ig_{ms}\,m\,q \nonumber\\
       &\quad + \Omega + \sqrt{2\kappa_m}\,m^{\rm in}, \\
\dot q &= \omega_s\,p, \\
\dot p &= -\omega_s\,q - \gamma_s\,p - g_{ms}\,m^{\dagger}m + \eta.
\end{align}
The quantum Langevin equations are nonlinear due to the magnetostrictive interaction term. Since the magnon mode is strongly driven, the cavity, auxiliary cavity, magnon, and mechanical modes evolve around well-defined steady-state operating points. The dynamics can therefore be described by considering small quantum fluctuations about these steady-state values. To this end, each operator is decomposed as

\begin{equation}
O=O_s+\delta O,
\qquad
O=c,a,m,
\end{equation}
together with \begin{equation}
q=q_s+\delta q,
\qquad
p=p_s+\delta p.
\end{equation}
Here,  $O_s=\langle O\rangle$ denotes the steady-state amplitude and
$\delta O$ represents the corresponding quantum fluctuation.
\ha{ Keeping only terms linear in the fluctuations yields Gaussian dynamics. The static mechanical displacement modifies the magnon detuning according to 
\begin{equation}
    \widetilde{\Delta}_m
    =
    \Delta_m+g_{ms}q_s,
\end{equation}
while the linearized magnomechanical interaction is characterized
by the effective coupling
\begin{equation}
G_{ms}
=
\sqrt{2}\,g_{ms}m_s,
\end{equation}
where the phase of the coherent magnon amplitude is chosen such
that $m_s$ is real.
}

Substituting these expressions into the quantum Langevin equations and
neglecting second-order fluctuation terms yields a linear set of equations
governing the evolution of the fluctuations.
For convenience, the bosonic fluctuation operators are expressed in terms
of their amplitude and phase quadratures,
\begin{equation}
\delta X_j=\frac{\delta j+\delta j^\dagger}{\sqrt2},
\qquad
\delta Y_j=\frac{i(\delta j^\dagger-\delta j)}{\sqrt2},
\end{equation}
with $j=c,a,m,$ which satisfy the canonical commutation relations. 
Using these quadratures, the state of the linearized system is completely specified by the fluctuation vector
\begin{equation}
\mathbf{u}=
(
\delta X_c,
\delta Y_c,
\delta X_a,
\delta Y_a,
\delta X_m,
\delta Y_m,
\delta q,
\delta p
)^T,
\end{equation}

whose dynamics can be written in the compact form

\begin{equation}
\dot{\mathbf{u}}(t)=A_k\,\mathbf{u}(t)+\mathbf{n}(t),
\end{equation}
where $n(t)$ is the Gaussian noise vector and the drift matrix is

\begin{widetext}
\begin{equation}
A_k=
\begin{pmatrix}
-\kappa_c & \Delta_c & 0 & J & 0 & G_{mc} & 0 & 0 \\
-\Delta_c & -\kappa_c & -J & 0 & -G_{mc} & 0 & 0 & 0 \\
0 & J & -\kappa_a & \Delta_a^{(k)} & 0 & 0 & 0 & 0 \\
-J & 0 & -\Delta_a^{(k)} & -\kappa_a & 0 & 0 & 0 & 0 \\
0 & G_{mc} & 0 & 0 & -\kappa_m & \widetilde{\Delta}_m & 0 & 0 \\
-G_{mc} & 0 & 0 & 0 & -\widetilde{\Delta}_m & -\kappa_m & -G_{ms} & 0 \\
0 & 0 & 0 & 0 & 0 & 0 & 0 & \omega_s \\
0 & 0 & 0 & 0 & -G_{ms} & 0 & -\omega_s & -\gamma_s
\end{pmatrix}.
\label{eq:drift_matrix}
\end{equation}
\end{widetext}

Since the linearized dynamics preserve Gaussianity, the reservoir state is completely characterized by the covariance matrix

\begin{equation}
(V_{\ha{k}})_{ij}
=
\frac12
\left\langle
u_i u_j
+
u_j u_i
\right\rangle.
\end{equation}

The covariance matrix satisfies the continuous-time Lyapunov equation

\begin{equation}
\dot V
=
A_kV_k
+
V_kA_k^T
+
D,
\label{eq:lyap}
\end{equation}

where $D$ is the diffusion matrix describing the quantum and thermal
fluctuations associated with the dissipative channels of the cavity,
auxiliary cavity, magnon, and mechanical modes. Assuming independent
Markovian environments, the diffusion matrix is diagonal and can be written as
\begin{equation}
    \begin{aligned}
    D &=
    \mathrm{diag}
    [
    \kappa_c(2N_c+1),
    \kappa_c(2N_c+1),
    \kappa_a(2N_a+1),
    \\
    & 
    \kappa_a(2N_a+1),
    \kappa_m(2N_m+1),
    \kappa_m(2N_m+1),
    0,\\
    &
    \gamma_s(2N_s+1)
    ],
    \label{eq:diffusion}
    \end{aligned}
\end{equation}

where $N_j\left(\omega_j\right)=\left(\exp \left[\hbar \omega_j / K_B T\right]-1\right)^{-1}$ (with $j=\gl{a,}c, m, s$ ) represents the mean thermal occupation
numbers of the cavity, auxiliary cavity, magnon, and mechanical reservoirs,
respectively.

\section{Cavity magnomechanical quantum reservoir computing}\label{sec:III}
%\subsection{Input encoding}
\gl{Having established the linearized Gaussian dynamics of the hybrid system, we now
operate this platform as a quantum reservoir, using the covariance matrix as the
source of the reservoir features processed by a trained linear readout.} At each time step \ha{k}, the input $s_k$ is encoded into the auxiliary cavity through the detuning $\Delta_a^{(k)}$ (Eq. \ref{eq:encoding}). The encoded input is kept constant during an interval $\Delta t$, so that the drift matrix $A_k$ remains constant over this interval. The continuous covariance dynamics described by Eq. (\ref{eq:lyap}) are then propagated from one input step to the next using the Van Loan discretization method \cite{VanLoan1978}, yielding
\begin{equation}
V_{k+1}
=
\Phi_kV_k\Phi_k^T
+
Q_k,
\label{eq:propagation}
\end{equation}

where
\begin{equation}
\Phi_k=e^{A_k\Delta t}
\end{equation}

is the state-transition matrix and

\begin{equation}
Q_k
=
\int_0^{\Delta t}
e^{A_k\tau}
D
e^{A_k^T\tau}
\,d\tau
\end{equation}
is the accumulated noise covariance.

\gl{It is worth stressing where the nonlinearity of the reservoir comes from.
The fluctuation dynamics of Eq.~(\ref{eq:lyap}) is linear in $\mathbf{u}$,
and one might expect a Gaussian reservoir to implement only linear
transformations of its input. The input, however, does not enter as an
additive drive but through the drift matrix itself, via the detuning
$\Delta_a^{(k)}$ in Eq.~(\ref{eq:drift_matrix}). As a consequence, both the
state-transition matrix $\Phi_k=e^{A_k\Delta t}$ and the accumulated noise
covariance $Q_k$ are nonlinear functions of $s_k$, and the recursion of
Eq.~(\ref{eq:propagation}) generates products of such matrices across
successive time steps. The elements of $V_k$ are therefore nonlinear
functions of the whole input history, with cross terms coupling inputs at
different times. This mechanism, by which nonlinearity in the input--output
map emerges from the encoding even when the underlying dynamics is linear,
has been identified analytically for continuous-variable
reservoirs~\cite{Mujal2021b, Nokkala2021}, and it is what allows the present
platform to address nonlinear temporal tasks with a purely linear readout.}

The covariance matrix obtained after each propagation step defines the reservoir state associated with the corresponding input sample. The resulting reservoir states are represented by feature vectors $\mathbf{x}_k$ and assembled into the design matrix $X=[\mathbf{x}_1,\mathbf{x}_2,\hdots,
\mathbf{x}_{N}]^T$ where $N$ denotes the total number of input samples. The design matrix is then used to train the linear readout layer.

The reservoir parameters remain fixed throughout the learning process, and only the output weights are optimized. Given the target vector $Y=[y_1, y_2, \hdots,y_N]^T$, the readout weights are determined using ridge regression,
$ W_{\mathrm{out}} = \left(X^{T}X+\lambda I\right)^{-1}X^{T}Y,$ where $\lambda$ is the regularization parameter. 
The predicted outputs are then obtained as
$\hat{Y}=XW_{\mathrm{out}}.$
The performance is quantified using the normalized mean-square
error,

\begin{equation}
\mathrm{NMSE}
=
\frac{
\left\langle
(y_k-\hat y_k)^2
\right\rangle
}{
\operatorname{Var}(y_k)
}.
\label{eq:nmse}
\end{equation}

A smaller NMSE therefore corresponds to a more accurate
reconstruction of the target sequence.

\section{Results and Discussion}

In this section, we first investigate the computational capabilities of the proposed system as a quantum reservoir. The reservoir is evaluated on representative benchmark tasks to assess its ability to process temporal information and perform nonlinear computations. We then demonstrate how these computational properties can be exploited for probing an external signal encoded through the auxiliary-cavity detuning. \ha{We also analyze the influence of the main physical parameters on the reconstruction accuracy and assess its robustness against Gaussian detuning noise for different reservoir feature spaces.}

\subsection{Benchmark tasks}
A suitable reservoir must simultaneously exhibit fading memory, which enables information about previous inputs to persist over time, and nonlinear dynamics that allow complex temporal transformations to be performed. To evaluate these properties, we consider three benchmark tasks widely used in the quantum reservoir computing literature \cite{Fujii_2017}.

The first benchmark is the linear memory task,
\begin{equation}
y_k=s_{k-\tau},
\end{equation}
which evaluates the ability of the reservoir to reconstruct delayed inputs. The second benchmark is the nonlinear memory task,
\begin{equation}
y_k=\left(s_{k-\tau}\right)^2,
\end{equation}
which probes the nonlinear processing capability of the reservoir. Finally, we consider the parity-check benchmark \cite{NIPS2004_f8da71e5},
\begin{equation}
y_k=
\left(
\sum_{j=0}^{d-1}b_{k-j}
\right)
\bmod 2,
\end{equation}
where $b_k\in\{0,1\}$ denotes a binary input sequence. Since the parity order starts from $d=1$, we plot the parity-check results using the correspondence $d=\tau+1$ to facilitate comparison with the memory tasks on a common horizontal axis.

\begin{figure}[h]
    \centering
    \includegraphics[width=1\linewidth]{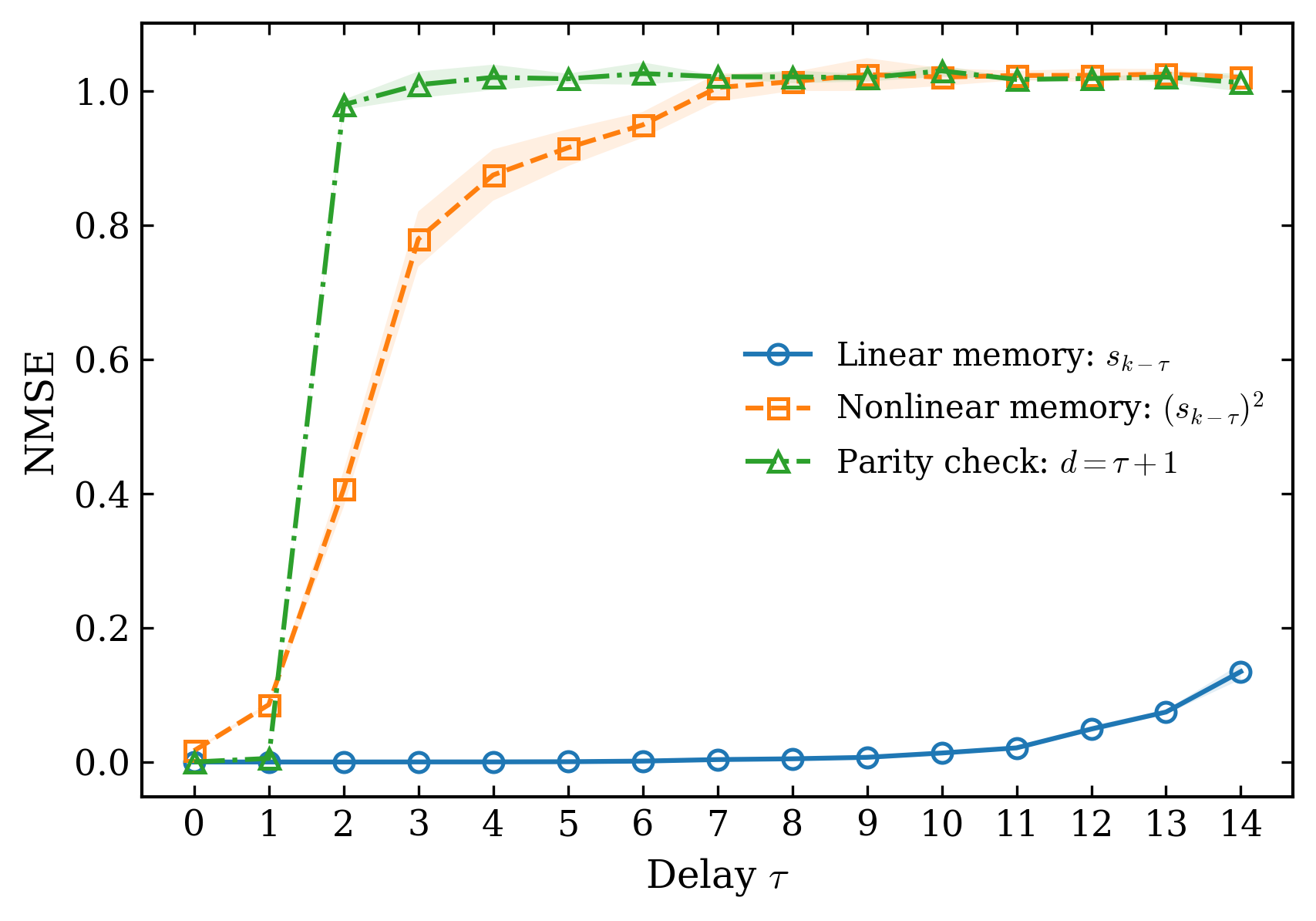}
    \caption{
Performance of the proposed extended hybrid cavity
magnomechanical quantum reservoir on three benchmark tasks.
The NMSE is shown as a function of the delay $\tau$ for the
linear-memory task, $y_k=s_{k-\tau}$ (blue circles), and the
nonlinear-memory task, $y_k=(s_{k-\tau})^2$ (orange squares).
For the parity-check task (green triangles), the parity order is
represented as $d=\tau+1$ to allow the three benchmarks to be
displayed on a common horizontal axis.
}
    \label{fig:benchmark}
\end{figure}

\begin{figure}[h]
    \centering
    \includegraphics[width=1\linewidth]{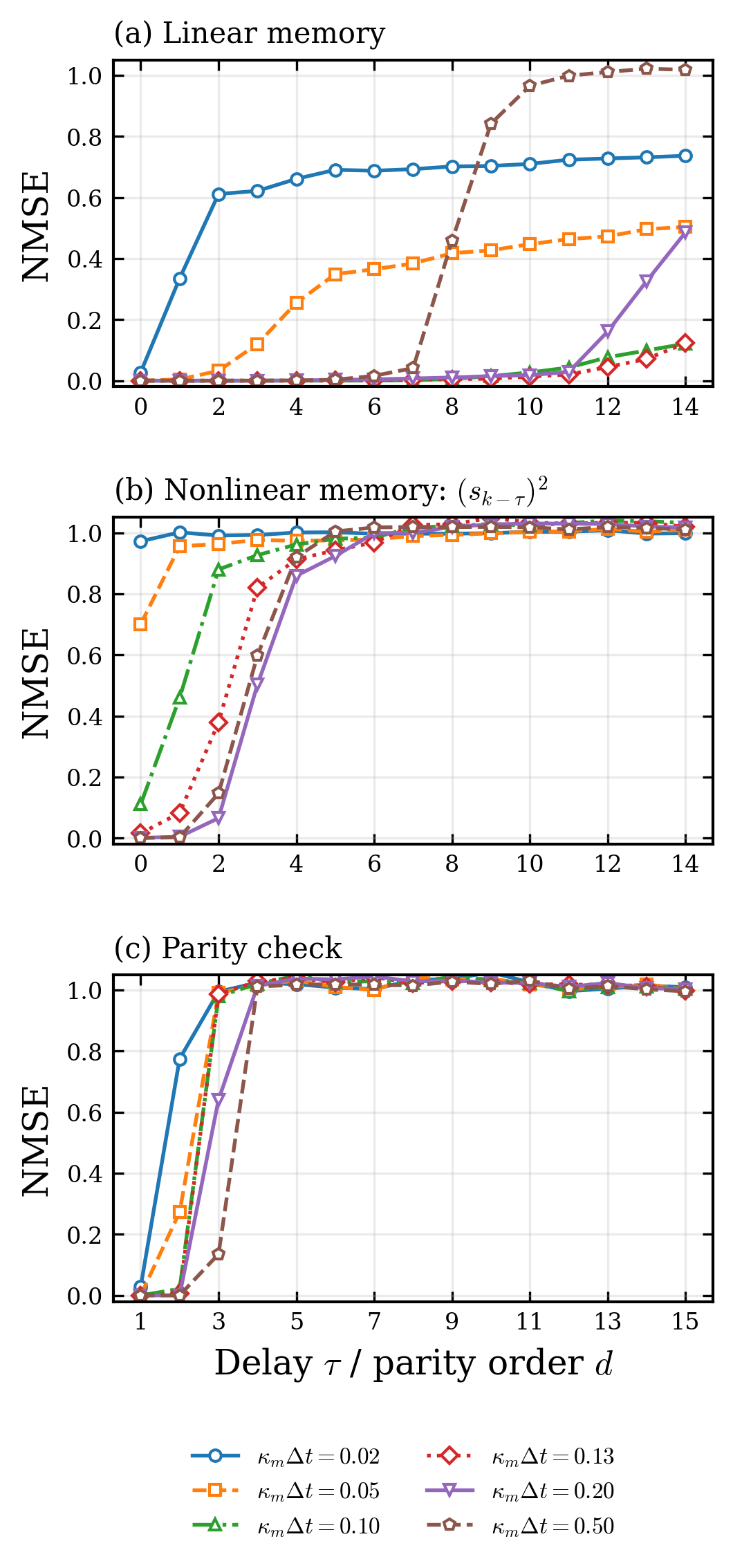}
    \caption{Effect of $\kappa_m\Delta t$ on the benchmark performance of the proposed reservoir. (a) Linear memory task, (b) nonlinear memory task with target $y_k=(s_{k-\tau})^2$, and (c) parity-check task.}
    \label{benchmark_dt}
\end{figure}
    
Fig.~\ref{fig:benchmark} compares the performance of the
reservoir on the three benchmark tasks. A pronounced difference is observed between the linear-memory and nonlinear-processing tasks. 
For the linear-memory benchmark, the NMSE remains very small over a
broad range of delays and increases appreciably only for the largest values of $\tau$. This behavior indicates that information about past inputs is efficiently retained by the reservoir, demonstrating a
relatively long fading memory.

\ha{The nonlinear memory task shows a faster increase of the reconstruction error with the delay, since the reservoir must both retain the delayed input and provide a nonlinear representation accessible to the linear readout. This distinction is particularly relevant for the present reservoir, whose
fluctuation dynamics is linear. As discussed in Sec.~III, nonlinearity does
not originate from the dynamics but from the encoding: since $s_k$ modulates
the drift matrix, the covariance elements are nonlinear functions of the input
history, and the nonlinearity accessible to the readout depends on the
encoding scheme and on the measured observables~\cite{Mujal2021b, Nokkala2021}.
The faster growth of the error in the nonlinear-memory task thus indicates
that, for the present detuning encoding and covariance features, the
nonlinear content of the reservoir response is weaker than its linear
memory, rather than an intrinsic limitation of the platform.} 

\ha{The parity-check task constitutes an even more stringent benchmark, since the target depends nonlinearly on several previous inputs. The NMSE rapidly approaches unity as the parity order increases, showing that the simultaneous retention and nonlinear combination of multiple
past inputs represents the most demanding task considered here.}

\ha{Taken together, these benchmarks reveal a reservoir whose dominant
computational resource is temporal memory, accompanied by a more
limited nonlinear processing capability. This imbalance motivates the
analysis of the reservoir evolution time presented below, where we
investigate how the dynamical operating point can modify the balance
between memory retention and nonlinear processing.}

Fig.~\ref{benchmark_dt} illustrates how the reservoir evolution time $\kappa_m\Delta t$ affects the computational performance of the proposed reservoir. For the linear memory task, smaller values of $\kappa_m\Delta t$ preserve information over longer delays, whereas increasing the evolution interval leads to a faster degradation of the memory. In contrast, the nonlinear benchmark shows that intermediate values of $\kappa_m\Delta t$ improve nonlinear processing, highlighting the role of reservoir dynamics in generating nonlinear transformations. The parity-check task exhibits a weaker dependence on $\kappa_m\Delta t$, with all tested values achieving comparable performance for low parity orders, while the performance progressively deteriorates as the task requires information from a larger number of past inputs. 

\gl{Overall, these results reveal a trade-off between temporal memory and
nonlinear processing, controlled by the evolution interval between
successive input injections. Such a trade-off is expected on general
grounds: the information processing capacity of a reservoir, which
quantifies its ability to reconstruct linear and nonlinear functions of past
inputs, is bounded by the number of linearly independent features available
to the readout, so that memory and nonlinear processing compete for a finite
total capacity~\cite{Dambre2012, MartinezPena2023}. The evolution time
$\kappa_m\Delta t$ thus redistributes this capacity between the two without
changing the feature space. In the remainder of the analysis, we adopt
$\kappa_m\Delta t = 0.13$, which provides good memory retention while
maintaining accurate nonlinear processing at short delays.}

\subsection{Reservoir feature-space analysis}
In this section, we will examine how the choice of reservoir observables influences the information extracted from the reservoir state. Although the Gaussian reservoir is fully characterized by its covariance matrix, it is not evident whether all covariance elements are required or whether a reduced set of observables provides comparable performance. To address this question, we compare five feature spaces constructed from the covariance matrix: the auxiliary cavity, microwave cavity, and magnon mode individually, the local quadrature fluctuations of all modes, and the complete set of observables. For all cases, the reservoir dynamics, input sequence, and linear readout remain unchanged, ensuring that the comparison isolates only the effect of the selected observables. 

\begin{figure}[h]
    \centering
    \includegraphics[width=1\linewidth]{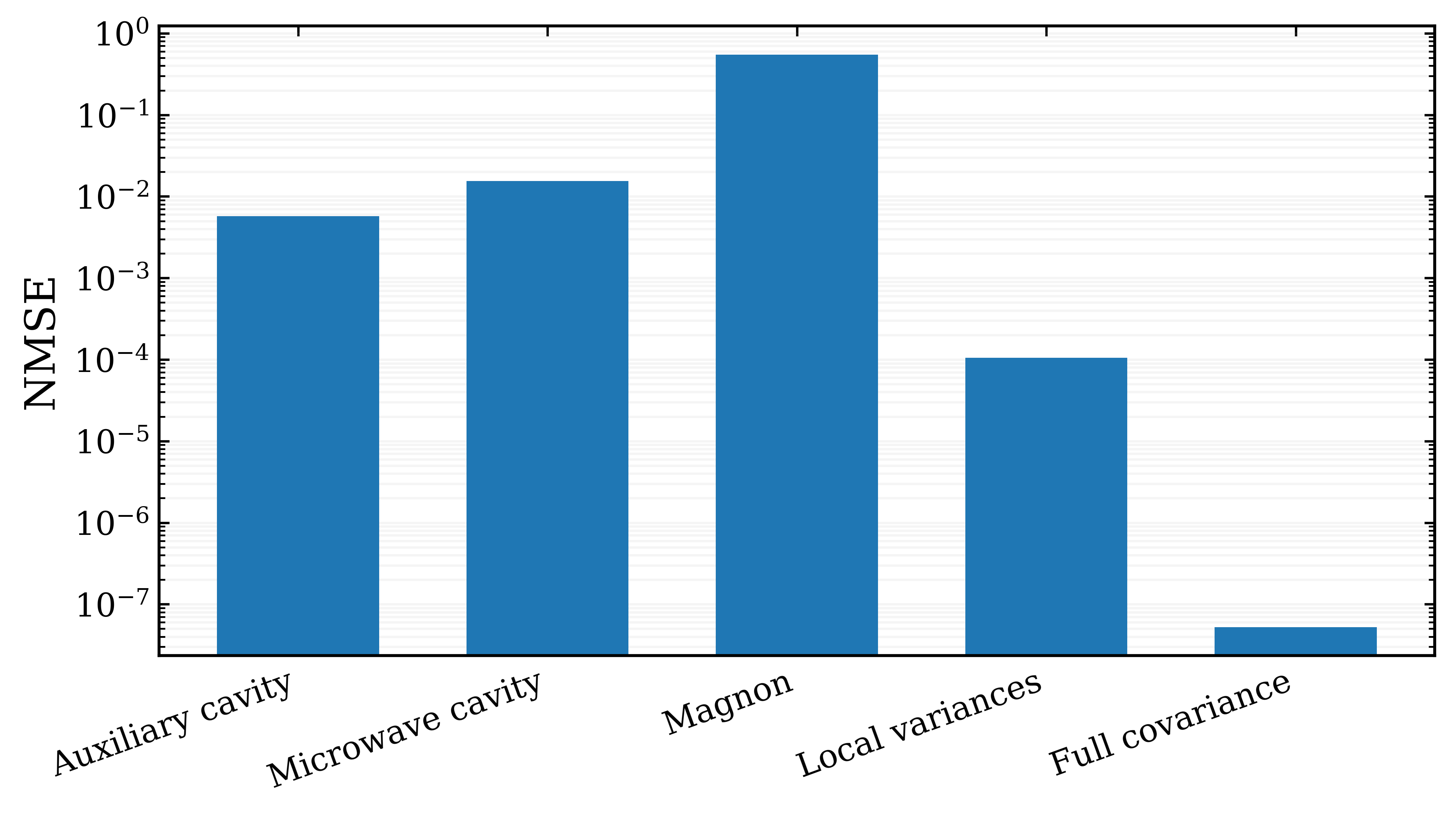}
    \caption{Comparison of probing performance across different reservoir feature sets.}
    \label{fig:features}
\end{figure}

As shown in Fig.~\ref{fig:features}, measurements restricted to individual subsystems provide only limited information, while using the local quadrature fluctuations of all modes significantly improves the reconstruction accuracy. The best performance is achieved when the complete set of observables is employed, demonstrating that correlations among the different modes play an important role in the reservoir dynamics. Therefore, the complete feature space is adopted throughout the remainder of this work.

\subsection{Signal Probing with the Extended Magnomechanical Quantum Reservoir }%EOQR)}

%\subsection{Reservoir-Based Quantum Sensing}

Having established the computational capability of the proposed reservoir through the benchmark tasks, we now investigate its application to signal probing. Specifically, we consider the reconstruction of an external classical signal encoded in the auxiliary-cavity detuning. At each time step, the input signal is injected according to Eq.~\ref{eq:encoding}, where $s_k$ denotes the
signal to be reconstructed. The objective is therefore to reconstruct the encoded signal directly from the reservoir response.

To analyze the influence of the main system parameters on the reconstruction accuracy, we vary them relative to the reference operating point specified in Table~\ref{tab:parameters}. We denote by $J_0$, $G_{mc,0}$, $G_{ms,0}$, and $\epsilon_{\Delta,0}$ the reference values of the auxiliary-cavity coupling, cavity-magnon coupling, effective magnomechanical coupling, and encoding strength, respectively. The normalized ratios $J/J_0$, $G_{mc}/G_{mc,0}$, $G_{ms}/G_{ms,0}$, and $\epsilon_\Delta/\epsilon_{\Delta,0}$ therefore provide a common dimensionless scale for comparing the sensitivity of the probing performance to variations in parameters with different characteristic magnitudes, with unity corresponding to the reference value of each parameter.

\begin{figure}[h]
    \centering
    \includegraphics[width=1\linewidth]{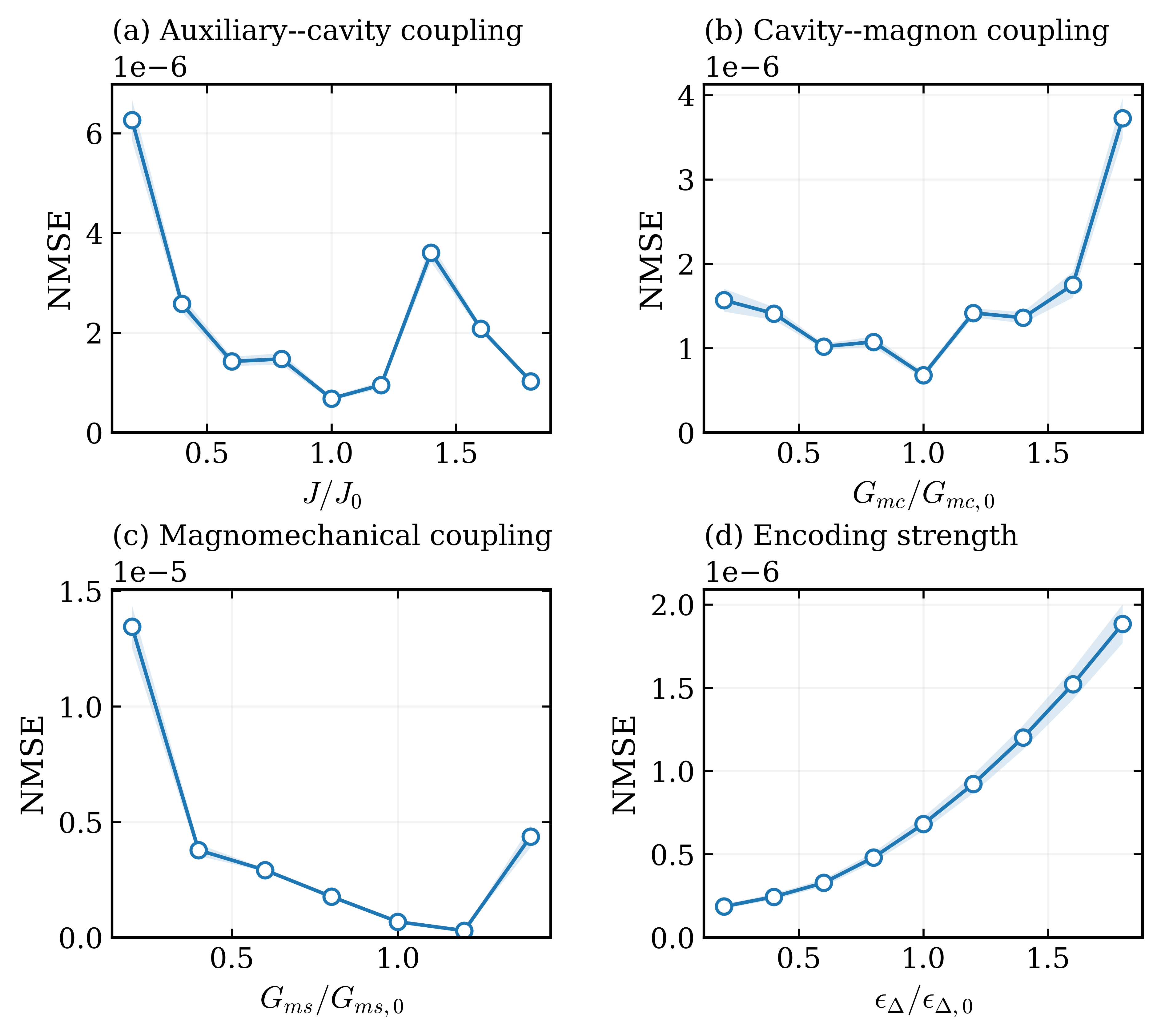}
    \caption{Probing performance as a function of the main reservoir parameters:
(a) $J$, (b) $G_{mc}$, (c) $G_{ms}$, and (d) $\epsilon_\Delta$.} 
    \label{fig: parameters sensing}
\end{figure}

We first examine how the reconstruction accuracy depends on the parameters governing the coupled reservoir dynamics and the input encoding. Fig.~\ref{fig: parameters sensing} shows the NMSE as a function of the auxiliary-cavity coupling $J$, the cavity-magnon coupling $G_{mc}$, the effective magnomechanical coupling $G_{ms}$,
and the encoding amplitude $\epsilon_\Delta$. Each parameter is varied independently around its reference value while the remaining parameters are kept fixed.

The probing performance exhibits a distinct dependence on the different physical parameters. The variation with $J$ is nonmonotonic, with the reconstruction error reaching its lowest values at an intermediate coupling strength. Since the signal is encoded in the auxiliary cavity, $J$ controls
the first stage through which the input perturbation is transferred to the rest of the reservoir. A weak coupling limits this transfer, whereas increasing
$J$ does not continuously improve the reconstruction, indicating that the reconstruction accuracy depends on a suitable dynamical balance between the auxiliary cavity and the internal reservoir modes.

A similar nonmonotonic behavior is observed for $G_{mc}$, with the lowest NMSE obtained close to the reference cavity--magnon coupling. Together with the dependence on $J$, this shows that stronger intermode coupling is not by
itself sufficient to improve the reconstruction accuracy. Rather, the coupled modes must operate in a regime in which the input-induced perturbation produces a reservoir response that can be efficiently decoded. The dependence on $G_{ms}$ further
supports this picture: increasing the effective magnomechanical coupling substantially improves the reconstruction up to a favorable range, while a
further increase leads to a deterioration of the performance.

Interestingly, the encoding amplitude $\epsilon_\Delta$ exhibits a different
trend, with the NMSE increasing as the modulation strength is increased.
Since $\epsilon_\Delta$ determines the range over which the input modulates the auxiliary-cavity detuning, a larger value does not simply amplify the information supplied to the reservoir, but also changes the dynamical regime explored during the input sequence. The observed behavior therefore shows that stronger input encoding does not necessarily produce a more accurately decodable reservoir response. Overall, these results indicate that probing
performance is governed by the interplay between the encoding strength and the internal couplings of the hybrid reservoir, rather than by maximizing any single interaction strength.

To assess the robustness of the probing protocol against imperfections in the input encoding, we introduce Gaussian fluctuations in the auxiliary-cavity detuning. The noisy encoding is described by
\begin{equation}
    \Delta_a^{(k)}
    =
    \Delta_{a,0}
    + \epsilon_{\Delta} s_k
    + \sigma_{\Delta} \xi_k,
\end{equation}
where $\xi_k \sim \mathcal{N}(0,1)$ is an independent Gaussian random variable and $\sigma_{\Delta}$ determines the noise amplitude. By varying $\sigma_{\Delta}$, we investigate how increasing detuning noise affects the reconstruction of the input signal. We also compare different sets of reservoir features to examine how the information extracted from the reservoir influences the robustness of the signal probing.
\begin{figure}[h]
    \centering
    \includegraphics[width=1.0\linewidth]{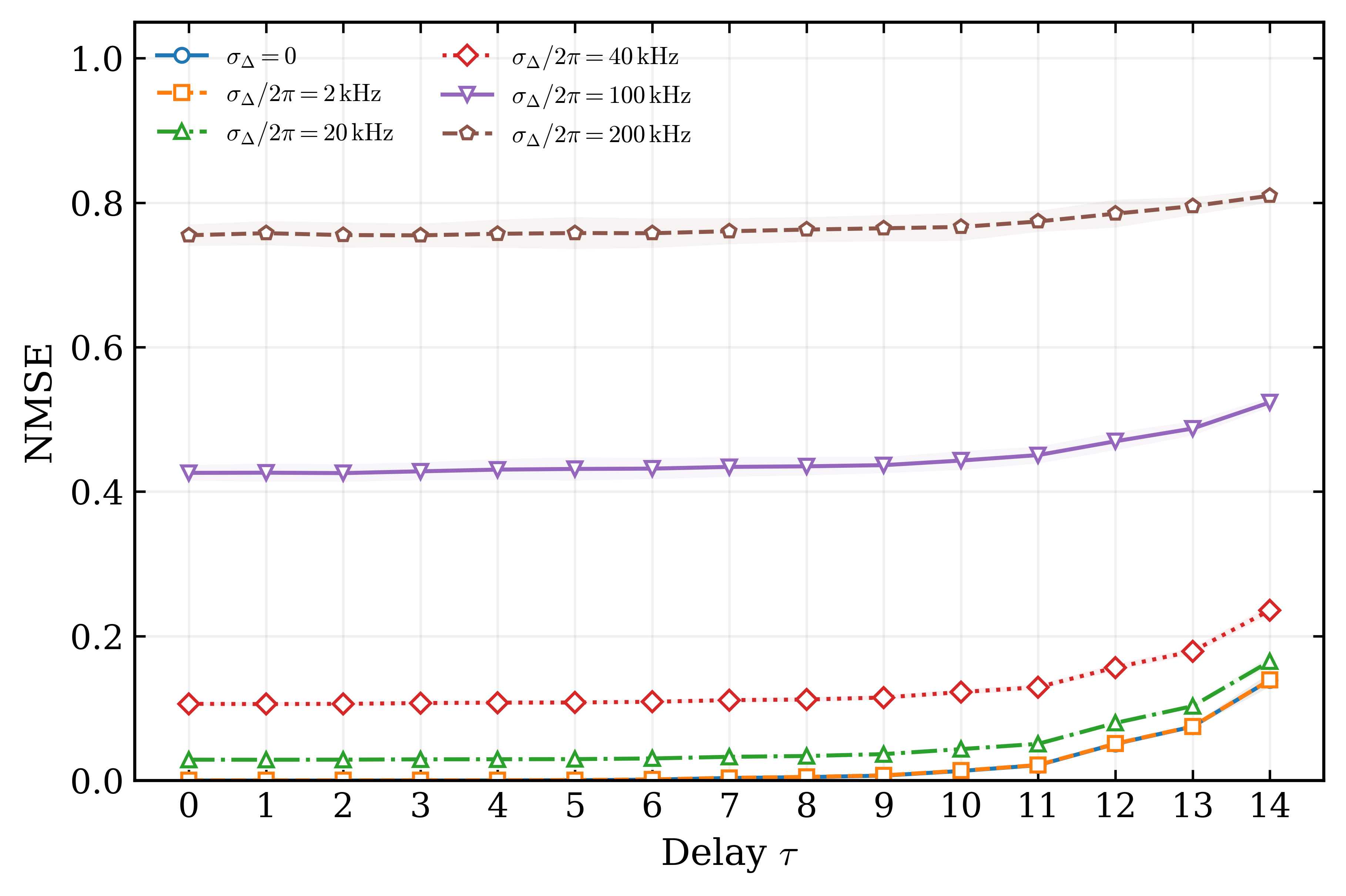}
    \caption{NMSE as a function of the delay $\tau$ for different amplitudes of Gaussian detuning noise $\sigma_\Delta$.}
    \label{fig:gaussian_noise}
\end{figure}
Fig.~\ref{fig:gaussian_noise} shows the combined effect of detuning noise and temporal delay on the reconstruction performance. In the absence of noise, the NMSE remains very small over a broad range of delays and increases only at larger values of $\tau$, consistent with the linear-memory behavior discussed above. As $\sigma_\Delta$ increases, the reconstruction error increases throughout the delay range. This degradation can be understood from the encoding mechanism itself: the Gaussian fluctuations perturb the same auxiliary-cavity detuning through which the signal is injected, thereby reducing the distinguishability of the signal-induced modulation in the subsequent reservoir response. For each noise level, the NMSE remains relatively weakly dependent on $\tau$ at short delays before increasing at longer delays, where the finite memory of the reservoir introduces an additional limitation. Interestingly, the two effects appear largely distinct: detuning noise primarily sets the overall error level, whereas the delay-dependent increase reflects the progressive loss of stored temporal information. Consequently, even though the reservoir retains a relatively long linear memory, this memory cannot compensate for information that has already been degraded during the encoding process. The results therefore show that delayed signal reconstruction is determined by both the quality of the input encoding and the ability of the reservoir to retain the encoded information over time. 
\begin{figure}[h]
    \centering
    \includegraphics[width=1.\linewidth]{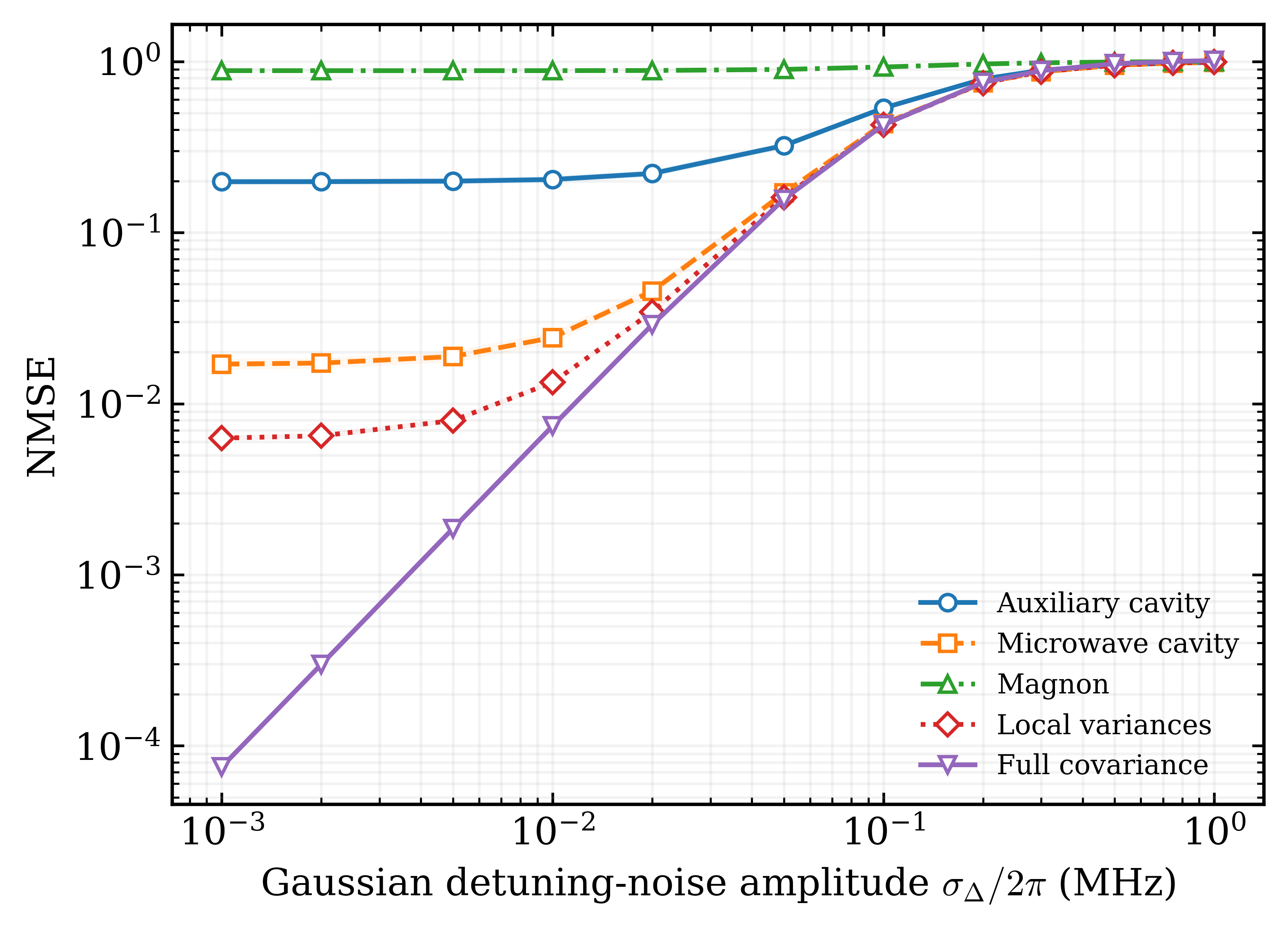}
    \caption{Probing performance under Gaussian detuning noise for different reservoir feature spaces}
    \label{fig:noise_delay}
\end{figure}
The robustness to detuning noise depends strongly on the reservoir features
used for reconstruction, as shown in Fig.~\ref{fig:noise_delay}. In the
low-noise regime, a clear hierarchy is observed among the different feature
sets. The full covariance matrix provides the lowest NMSE, outperforming the local variances and the features restricted to individual subsystems by several orders of magnitude. In particular, the poor performance obtained from the magnon features alone indicates that the information about the encoded signal is not equally accessible from all parts of the hybrid reservoir. The substantial improvement obtained when the full covariance matrix is used highlights the importance of the correlations established between the coupled modes for signal reconstruction.

As the detuning noise increases, the reconstruction performance progressively deteriorates for all feature sets, and the differences between them become
smaller. Interestingly, the advantage provided by the full covariance matrix
is most pronounced in the low-noise regime and gradually disappears as the
noise becomes comparable to the characteristic scale of the input modulation. At sufficiently large $\sigma_\Delta$, all curves approach
$\mathrm{NMSE}\simeq1$, indicating that the reconstruction becomes ineffective regardless of the information extracted from the reservoir. In this regime, the performance is therefore limited primarily by the corruption of the
encoded input rather than by the choice of reservoir features.

\subsection{Finite measurement statistics}
\label{sec:finite_measurements}

The results discussed above assume ideal access to the reservoir
covariance matrix $V_k$. In practice, however, its elements must be
estimated from a finite number of measurements. To account for this
statistical uncertainty, we consider $N_m$ independent realizations
of the reservoir quadratures at each time step. Since the fluctuation
quadratures have zero mean, the corresponding finite-sample covariance
matrix is estimated as
\begin{equation}
    \widehat V_k =
    \frac{1}{N_m}
    \sum_{r=1}^{N_m}
    \mathbf{u}_k^{(r)}
    \mathbf{u}_k^{(r)T},
    \label{eq:sample_covariance}
\end{equation}
where $\mathbf{u}_k^{(r)}$ denotes the $r$th realization of the quadrature vector. For a Gaussian state, the finite-sample covariance matrix $\widehat V_k$ follows Wishart statistics, with
fluctuations around $V_k$ decreasing as the number of measurements $N_m$ increases~\cite{wishart1928generalised}. In the simulations,
$\widehat V_k$ is used only to construct the features supplied to the readout, while the ideal covariance matrix $V_k$ is retained for the reservoir evolution. Thus, finite measurement statistics affect the
estimated features without altering the underlying reservoir dynamics.

\begin{figure}
    \centering
    \includegraphics[width=1\linewidth]{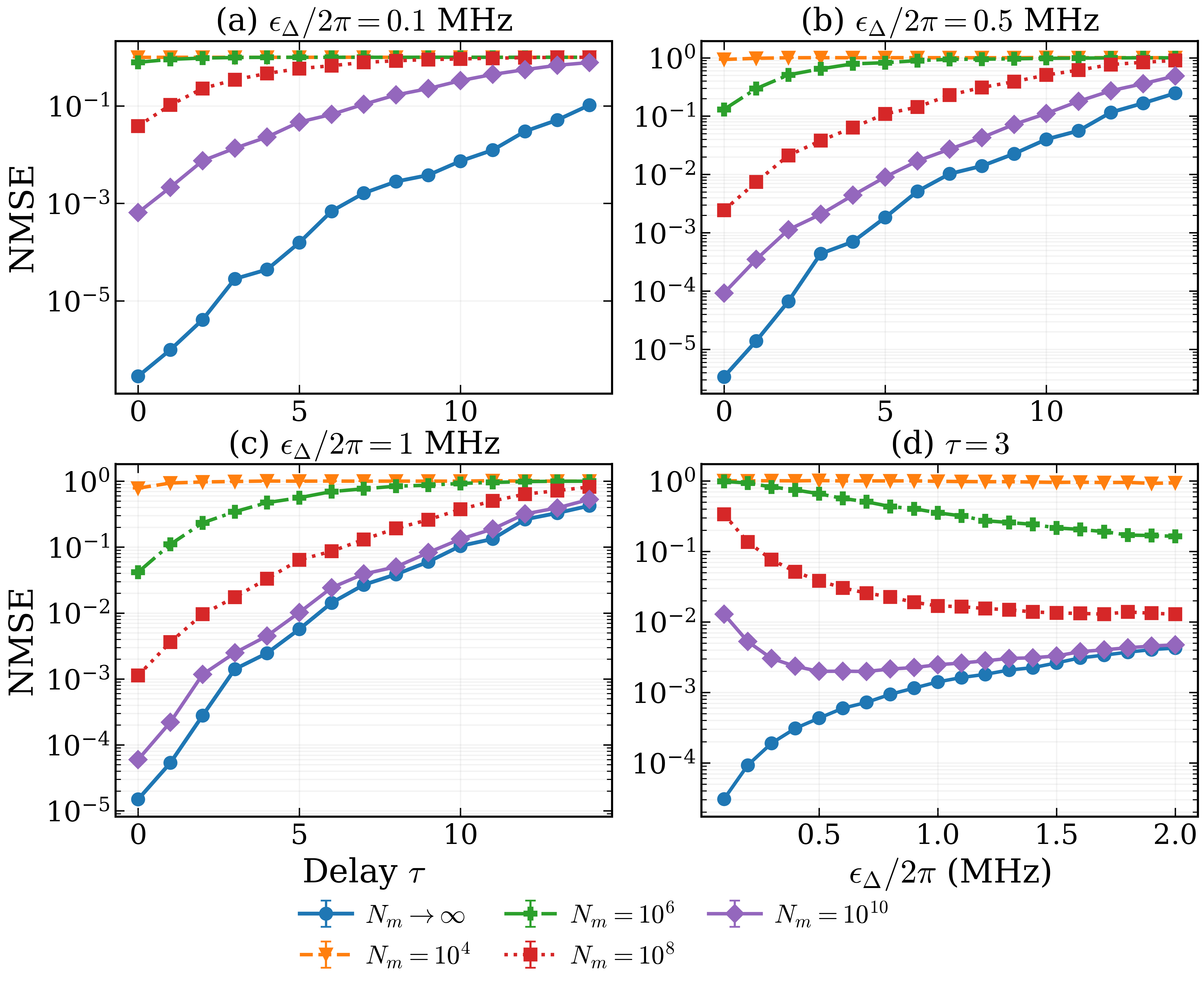}
    \caption{Effect of finite measurement statistics on the delayed signal probing performance. Panels (a)-(c) show the NMSE as a function of the delay $\tau$ for three input-encoding strengths, $\epsilon_\Delta/2\pi=0.1$, $0.5$, and $1$ MHz, respectively. Panel (d) shows the NMSE as a function of $\epsilon_\Delta$ at fixed delay $\tau=3$. Results are shown for $N_m=10^4$, $10^6$, $10^8$, and $10^{10}$ measurements and compared with the ideal $N_m\rightarrow\infty$ limit.}
    \label{fig:finite_measurements}
\end{figure}

Fig.~\ref{fig:finite_measurements} shows the effect of finite measurement statistics on the delayed reconstruction task. In Figs.~\ref{fig:finite_measurements}(a)-(c), the NMSE is shown as a function of the delay $\tau$ for three encoding strengths,
$\epsilon_\Delta/2\pi=0.1$, $0.5$, and $1$ MHz, respectively. For each case, the finite-measurement results are compared with the ideal $N_m\rightarrow\infty$ limit. As expected, increasing $N_m$ progressively improves the reconstruction and brings the performance
closer to the ideal result. For small $N_m$, the covariance-estimation uncertainty strongly limits the accessible memory, whereas for larger $N_m$ the delayed reconstruction is recovered over a broader range
of $\tau$.

The influence of the encoding strength is further examined in
Fig.~\ref{fig:finite_measurements}(d), where the delay is fixed at $\tau=3$
and $\epsilon_\Delta$ is varied. The ideal and finite-measurement cases
display markedly different behaviors. In the ideal limit, where the covariance
matrix is known with infinite precision, even a weak input modulation can be
resolved. Increasing $\epsilon_\Delta$ then makes the map from the input to the
covariance elements more strongly nonlinear, as discussed in Sec.~\ref{sec:III}, and
therefore less accessible to a linear readout, which progressively increases
the NMSE.

For finite $N_m$, however, an additional effect arises from the statistical uncertainty in the estimated covariance matrix. When $\epsilon_\Delta$ is too small, the input-induced variations of the reservoir covariance can be masked by finite-sampling fluctuations.
Increasing $\epsilon_\Delta$ enhances these variations relative to the statistical uncertainty and therefore improves the reconstruction. This competition creates an intermediate regime where the encoding is strong enough to be resolved without excessively degrading the reservoir's intrinsic memory. This behavior is
clearly visible for $N_m=10^{10}$, for which the NMSE reaches a minimum at an intermediate value of $\epsilon_\Delta$ before increasing again. For smaller $N_m$, the minimum is not reached over the range of encoding strengths considered here, suggesting that it
may occur at larger $\epsilon_\Delta$.

\section{Conclusions}
We have proposed and analyzed a hybrid cavity magnomechanical system as a
Gaussian quantum reservoir. The platform combines microwave-cavity, magnon,
and mechanical degrees of freedom, and is extended with an auxiliary cavity
that serves as an input port: information is injected by modulating its
detuning, so that the intrinsic cavity--magnon--mechanical interactions remain
untouched. In the linearized regime the dynamics is Gaussian, and the covariance
matrix of the quadrature fluctuations, propagated exactly from one input step
to the next, provides the reservoir features processed by a linear readout.

The computational properties of the reservoir were first investigated
using linear-memory, nonlinear-memory, and parity-check benchmarks. Standard benchmarks show that the dominant computational resource of the
reservoir is temporal memory, with a relatively long fading memory in the
linear-memory task, accompanied by a more limited but tunable nonlinear
processing capability. The evolution time of the reservoir $\kappa_m\Delta t$ acts as a control parameter for the balance between memory retention and nonlinear
transformation, consistent with the known dependence of Gaussian reservoirs on
the input encoding and on the observables used for readout. The feature-space
analysis further shows that the intermode correlations contained in the full
covariance matrix substantially increase the information accessible to the
readout, well beyond what is provided by local quadrature fluctuations or by
individual subsystems.

We have also investigated the effect of finite measurement statistics
in the estimation of the reservoir covariance matrix. Finite sampling
introduces an additional limitation on the accessible temporal
information, which becomes progressively weaker as the number of
measurements increases. Importantly, the encoding strength plays
different roles in the ideal and finite-measurement regimes. While weak
encoding can provide high reconstruction accuracy when the covariance matrix is known with ideal precision, stronger encoding can improve the distinguishability of the signal-induced covariance variations against finite-sampling fluctuations. This competition leads to a trade-off between preserving the intrinsic reservoir memory and achieving sufficient statistical resolution.

Overall, our results show that hybrid cavity magnomechanical systems provide a
suitable dynamical architecture for Gaussian quantum reservoir computing. Their multimode dynamics supports temporal information processing, while correlations between the different degrees of freedom provide additional information that can be exploited by the readout. The signal-probing application further demonstrates how the same reservoir can process information encoded directly in a physical parameter of the system. The quantities reported here are reconstruction errors of a trained readout; the ultimate precision achievable by such a probing scheme is a distinct question, which we leave to future work. A further direction concerns the readout itself: we have assumed that the reservoir features are estimated from independent repetitions of the protocol, so that the measurements do not perturb the subsequent evolution. In an online setting, however, the reservoir is monitored continuously, and the measurement back-action modifies the very dynamics that performs the computation. Rather than a mere source of disturbance, such back-action can act as a controllable resource, providing effective dissipation and shaping memory and separability in monitored quantum reservoirs~\cite{Morgui2026}; extending the present analysis to continuous monitoring of the output modes is therefore a natural next step.

\begin{acknowledgments}
H.A. gratefully acknowledges the Institute for Cross-Disciplinary Physics and Complex Systems (IFISC, UIB-CSIC) for its hospitality during her research visit, during which part of this work was carried out. H.A. also thanks Khadija El Anouz for helpful discussions.
We acknowledge support from the Spanish State Research Agency, through the María de Maeztu project CEX2021-001164-M, funded by MICIU/AEI/10.13039/501100011033; through the CoQuSy project PID2022-140506NB-C21 and -C22 funded by MICIU/AEI/10.13039/50110001103 and by ERDF, EU; and through the QuantCom project CNS2024-154720, funded by MICIU/AEI/10.13039/501100011033 and co-funded by the European Union; the project is funded under the Quantera II program that has received funding from the EU’s H2020 research and innovation program under Grant Agreement No. 101017733, and from the Spanish State Research Agency (project QNet PCI2024-153410) funded by MICIU/AEI/10.13039/50110001103  and by ERDF, EU.
\end{acknowledgments}

%\appendix

%\section{Appendixes}

\bibliography{sample}% Produces the bibliography via BibTeX.
\end{document}